\documentclass{ws-p8-50x6-00}

\begin{document}

\title{Preliminary results of the Standard Model Higgs Boson search at LEP2 in 2000}

\author{M.\ Kopal}
\address{Dept.of Physics, Purdue University, West Lafayette,
IN 47907-1396, USA\\ E-mail: miroslav.kopal@cern.ch}


\maketitle

\abstracts{
A search for the Standard Model Higgs boson is performed using the data
collected by the four LEP experiments at center-of-mass energies between
202 GeV and 209 GeV. An average luminosity of 140 $pb^{-1}$ per experiment
has been analyzed. A 2.6 $\sigma$ excess is observed in the LEP-wide
combination for a Higgs boson mass hypothesis of around 114 GeV. All results
are preliminary.}

\section{Introduction}
The Standard Model has succesfully described properties of interactions between
fundamental particles up to the electroweak scale. There is still one particle
predicted by the theory yet to be discovered: the Higgs boson.
Since its mass cannot be predicted by the Standard Model alone, all possible
masses must be investigated. Consistency of the Standard Model parameters prefers
a low-mass Higgs with mass around 100 GeV. A Higgs mass below 107.9 GeV has
already been excluded at $95\%$ confidence level by previous LEP-wide
analyses\cite{r2}.

\section{Higgs production and decay}
At interaction energies available at LEP, the most dominant mechanism of Higgs
production is called ``Higgs-strahlung'' ($e^+e^- \rightarrow Z^* \rightarrow
hZ$), where the electron-positron annihilation produces an off-shell Z boson which
then radiates a Higgs boson.

The Higgs boson, once produced, immediately decays into other particles. The main
decay modes available to the Higgs at LEP energies are b-quark decay ($h \rightarrow
b\bar b$), $\sim 80\%$ of all decays,  and $\tau$-lepton decay ($h \rightarrow \tau
\bar\tau$), $\sim 12\%$ of all decays. With the combination of the Z boson decay
modes, the following four decay channels are analyzed, covering more than $90\%$ of
the total branching ratio.
\begin{itemize}
 \item four-jet channel, when the Z boson decays into quarks ($hZ \rightarrow b\bar
       bq\bar q$). The branching ratio is $\sim 60\%$.
 \item missing energy channel, when the Z boson decays into neutrino pairs ($hZ
       \rightarrow b\bar b\nu\bar\nu$). The branching ratio is $\sim 20\%$.
 \item lepton channel, when the Z boson decays into $e^+e^-$ or $\mu^+\mu^-$
       pairs ($hZ \rightarrow b\bar bl^+l^-$). The branching ration is $\sim 3\%$
       respectively.
 \item $\tau$ channel, with Z boson decays into $\tau^+\tau^-$ pairs ($hZ
       \rightarrow b\bar b\tau^+\tau^-$), the branching ratio is $\sim 3\%$) or;
       because of the similar topology, the Higgs boson deacays into
       $\tau^+\tau^-$ pairs and while the Z boson decays into quark pairs ($hZ
       \rightarrow \tau^+\tau^- q\bar q$), the branching ratio is $\sim 6\%$.
\end{itemize}

\section{Analysis}
The main aim of the analysis is to reduce Higgs-like Standard Model backgrounds
while retaining any possible Higgs signals. After neural network or cut based
selections are performed, the most discriminating variables are combined to form
the ``final discriminant'' which is a function of the Higgs mass hypothesis.

\begin{figure}[t]
\begin{center}
\epsfxsize=22.5pc 
\epsfbox{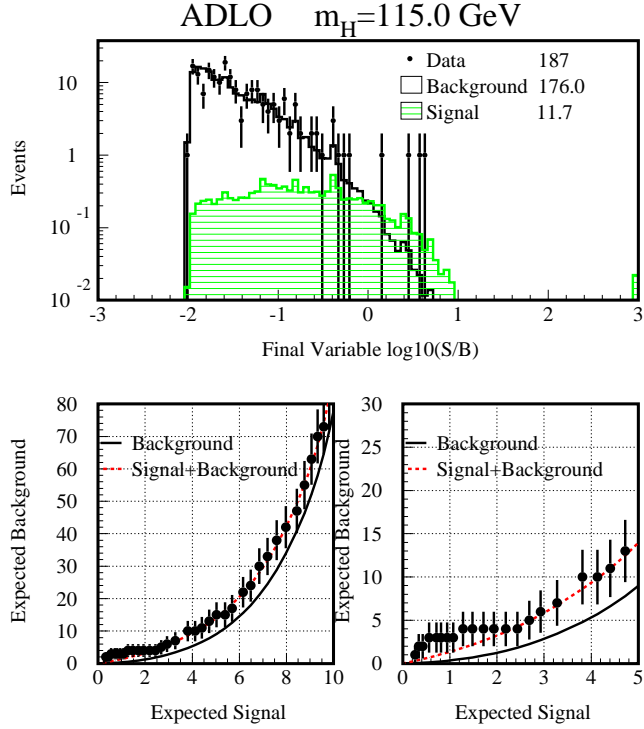} 
\vspace{-6pc}
\caption{The final discriminant variable for the Higgs mass hypothesis of 115 GeV.
\label{fig:f1}}
\end{center}
\end{figure}

Figure~\ref{fig:f1} shows the LEP-wide (ALEPH, DELPHI, L3 and OPAL - ``ADLO'')
final discriminant variable constructed for the Higgs mass hypothesis of 115 GeV.
In the upper plot, the white area is the background, the shaded area is the
Higgs signal and the dots are the observed data. The horizontal scale is the
logarithm base 10 of the signal to background ratio. Note the three right-most
events with high signal over background ratios.

The lower two plots show the integral of the upper plot from right to left. The
solid line is background only; the dashed line is signal plus background. Observed
data, indicated by dots, is more consistent with the signal plus background line.

An alternate way of determining whether a set of data is more consistent with
background rather than with signal plus background is to calculate the {\sl log
likelihood ratio}. The final variable is treated with Poisson statistics on a
bin-by-bin basis. The log likelihood ratio is then given by:
\begin{equation}
\ln Q = -S_{tot} + \sum_i n_i \ln (1 + \frac{S_i}{B_i})
\label{eq:eq1}
\end{equation}
where $n_i$ is the number of observed events in the i-th bin with $S_i$ expected
signal events and $B_i$ expected background events.

\begin{figure}[h]
\begin{center}
\epsfxsize=18.5pc 
\epsfbox{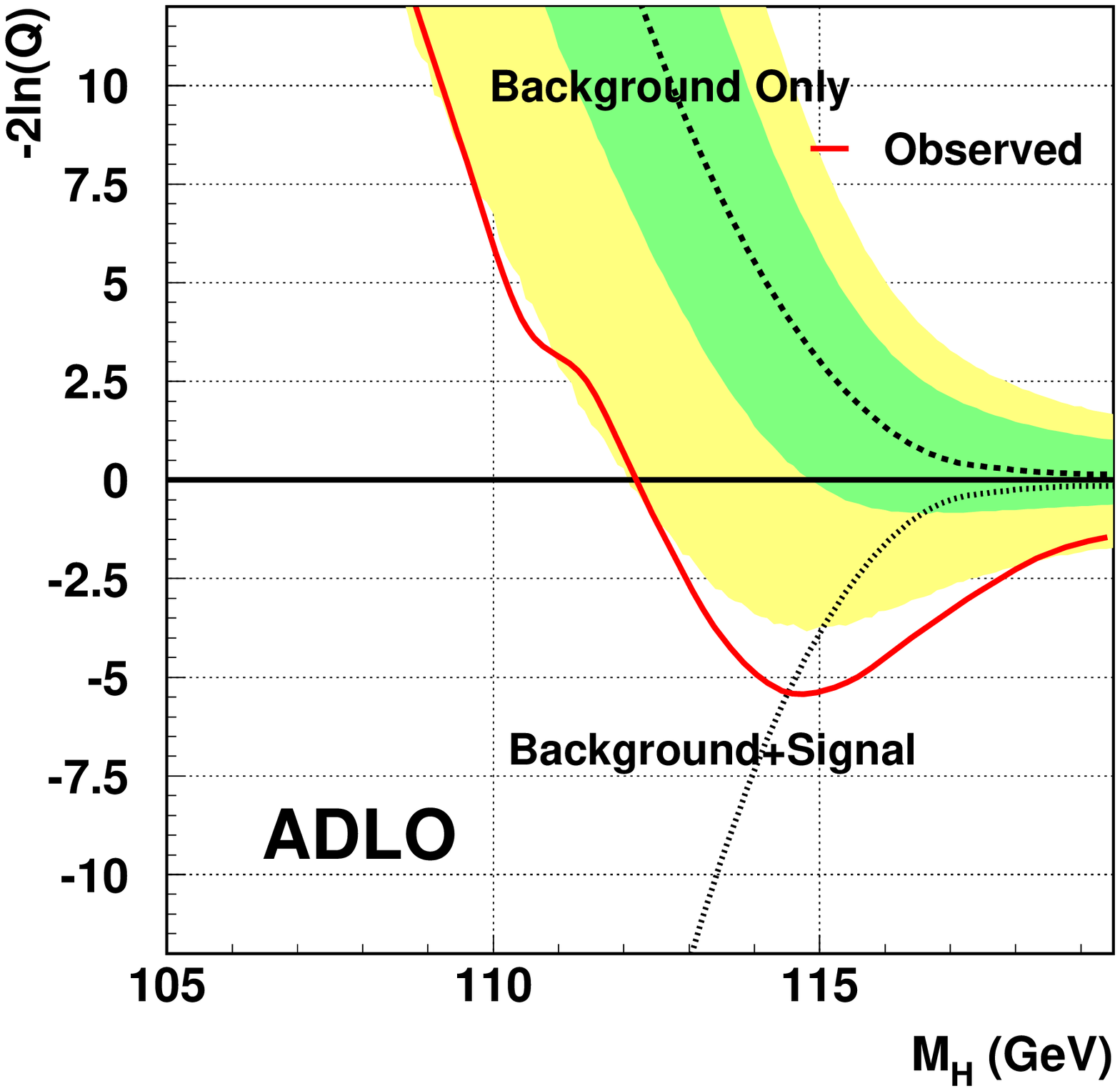} 
\vspace{-1pc}
\caption{Log Likelihood curve for LEP-wide data. \label{fig:f2}}
\end{center}
\end{figure}

The log likelihood plot is shown in Figure~\ref{fig:f2}.
The upper dashed line is the background only hypothesis ($n_i$ is the background
expectation), the lower dotted line is the signal plus background hypothesis ($n_i$
is signal plus background expectation). The solid line is the observed data.
The shaded region closest to the upper dashed line represents the area which is
less than 1 sigma away from background. The wider shaded region is 2 sigma away.

The data stays roughly at the 2 sigma border until about 114.5 GeV when it dips to
2.6 sigma off the background expectation. This dip is due mostly to Higgs-like
four-jet events seen by ALEPH and DELPHI.

\section{Conclusion}
With present statistics, Higgs observation cannot be confirmed or ruled out. The
LEP running has been extended until November 2 which will increase statistics. New
data could possibly exclude Higgs mass up to 114 GeV at $95\%$ confidence level or
increase the LEP-wide excess to 3 sigma. The results will be updated after the end
of 2000 LEP running. The results presented here are preliminary.


\end{document}